\documentclass[%
 reprint,
 amsmath,amssymb,
 aps,
]{revtex4-2}

\usepackage{graphicx}
\usepackage{dcolumn}
\usepackage{bm}

\begin{document}

\preprint{APS/123-QED}
\title{Emergence of rotational modes in nuclear fission}

\author{Bency John}
\altaffiliation[Also at ]{Homi Bhabha National Institute, Mumbai-400094, India.}
\email{bencyv.john@gmail.com}
\affiliation{%
	Nuclear Physics Division,\\ Bhabha Atomic Research Centre, Mumbai-400085, India
}%

\date{\today}

		\begin{abstract}
		Dinuclear systems that occur in  the post-saddle to  scission stage in nuclear  fission process are special transient formations. The diabatic evolution at this stage has been studied using the method of non-equilibrium thermodynamics. A novel explanation for the emergence of intrinsic rotational modes that occur in such  dinuclei has been developed by identifying these modes together as an open subsystem that exchanges matter and  energy with its environment. The environment consists of all  remaining degrees of freedom of the dinucleus, and a weak  coupling of the subsystem  to this environment facilitates  diabatic energy exchanges.  Under appropriate non-equilibrium conditions, the available energy is coupled to the work needed for the emergence of  self organizing rotational modes. This comes at the expense of increasing microscopic disorder in the form of intrinsic excitations in the forming fragments.  A simple formalism is presented such that the magnitude of relevant energy flow through the subsystem is obtained in terms of entropy production rate, entropy expulsion rate, and net rate of change of entropy in sub-units of  MeV~zs\textsuperscript{-1}~K\textsuperscript{-1}. 
	\end{abstract}
\maketitle	

%
%
\section{Introduction}
\label{int}
Spontaneous generation of intrinsic rotational modes   that cause ample gamma ray emissions from fission  fragments is a barely understood aspect in nuclear fission theory. In general, the nuclear rotation is a complicated interplay between the collective and intrinsic degrees of freedom. This situation is further enriched in the fission process since spontaneous  generation of intrinsic rotational modes occurs near the scission stage where the nucleus is already transformed into an excited  dinucleus under non-equilibrium conditions. While these modes carry no net angular momentum,  they can impart angular momenta to the individual fragments, and they occur relative to rigid rotation, if the latter is present in the dinucleus\cite{mor89,dosnp85,ran85}.  In induced fission reaction, very strong damping is present in the compound nuclear evolution just after the initial formation stage, and this strong damping phase continues  towards  the stages of   saddle and at times even beyond,  as long as the collective kinetic motion in elongation is slow. Thereafter, as the collective kinetic motions pick up speed,   the nature of damping changes to weak damping and this phase continues till final scission. The weak damping  is characterized by couplings between collective and intrinsic degrees and it allows elasto-plastic property for the nuclear matter \cite{nor84}.  The  weak damping  phase is  also known in literature as  non-adiabatic phase or as  diabatic   phase. In  fission, the collective pre-scission kinetic energy is largely accumulated during the weak damping  phase\cite{sim18}. This phase also allows generation of other collective degrees such as rotational and vibrational  modes that carry no net angular momentum.  Thus, overall, this is a very important phase of fission, however, diabatic descriptions  of the same are not as well developed as the adiabatic descriptions of the earlier phases. For this reason the nuclear scission process and the inherent dinuclear rotations are not well understood till this time.

As a  many body quantum system far from equilibrium, the dinucleus can manifest  quite different macroscopic degrees of freedom apart from  deformations.  Coherent and incoherent  excitation modes  can co-exist  in a dinucleus  with their characteristic growth  governed by the weak damping phase. The  free energy difference contributed  by the potential energy  release while moving from saddle to scission can set   stage for a diabatic environment that aids emergence of  ordered intrinsic rotational degrees from among multitude of other intrinsic excitations.  Thermodynamic sub-system concept that explain the emergence of intrinsic rotational degrees of freedom  in excited dinuclei is discussed in the present work, some probably for first time in fission theory. In the following, Sect.~\ref{met} describes the method for calculations. Physical environments of sliding,  rolling, and sticking  in the early phase of heavy ion collisions have similarity to  the environment of dinuclear rotations and hence the former is described in Sect.\ref{irmh} as a supporting  argument. The results are presented in Sect.~\ref{res} and a  discussion is given in  Sect.~\ref{dis}. Finally the summary and conclusions are given in Sect.~\ref{sum}. 

\section{Method}
\label{met}
\subsection{Intrinsic  rotational modes in fission}
\label{irm}
In the present work, the dinucleus at pre-scission stage  is treated as an overall closed thermodynamic system that encloses  an  open subsystem of  rotational degrees of freedom. This subsystem can exchange matter and energy with its surrounding heat bath that comprising of all remaining degrees of freedom.  Energy budget for the exchanges is accounted for by the pre-scission potential energy release.  For the intrinsic  rotational modes, the model  assumes their origination by nucleon pair transfers from the dinuclear neck region  aided by their characteristic couplings (dynamical pairs). This assumption may be justified since in the final phase, the volume from which balance nucleons retreat to either fragments is spatially confined to the neck volume. The nucleon retreats occur mostly pair wise, one nucleon each to each pre-fragment, and preferentially to the surface of pre-fragment (surface-peaked nature \cite{swi80}). The vector-sum of nucleon pair’s linear momenta will be approximately zero, and  the center of mass of the whole will be at rest. Although it is not necessary that these twin transfers occur simultaneously from the same location and they are of exactly opposite linear momentum, these qualities are to be maintained on an average, over a short time period, that satisfy the Heisenberg uncertainty relations. This mode of  retreat  of the nucleons accomplishes  a progressive necking-in and an eventual split that occurs when no nucleons are left in the neck. The average velocities with which these nucleons recede from the neck region to either side are close to  Fermi velocity of the nucleons. Grossly, the angular momentum imparted on a prefragment of radius vector $R^{'}$ by a single nucleon entering from neck with momentum $P_F$ (corresponding to its Fermi velocity) is  given approximately by $\frac{1}{2}R^{'}P_F\approx 4\hbar $ \cite{ran82}. Due to the linear and angular momenta conservations and other symmetries, near equal  angular momentum in same direction will be imparted on the other prefragment  by the partnering nucleon \cite{joh15}. The conservation of total angular momentum is achieved by the recoil  orbital angular momenta imparted at the center of mass of the dinucleus.  A part of the energy brought by the receding  pair will be converted thus as rotational energy. The rest energy is added to the heat bath.   Due to properties of the thermodynamic system bifurcation\cite{kar08}, at a certain  critical stage during this process, the angular momenta thus generated  reinforce themselves to be a part of the system collectivity. These gross properties need further qualifications so that different pattern of intrinsic modes generated are identified.

\subsection{Intrinsic  rotational modes in heavy ion collisions}
\label{irmh}
It is instructive to gather lessons from near symmetric heavy ion collisions  for further understanding the processes discussed above. 
Results from deep inelastic heavy ion collisions evidenced that the entrance channel orbital angular momentum($L$)  is dissipated gradually all the way down to the rigid rotation limit. This indicated the existence of an intrinsic mode that couples the entrance orbital angular momentum to the individual angular momenta of the collision partners($I_1$ and $I_2$), thereby implying  the existence of a wriggling type mode\cite{mor89}. Further more, from a macroscopic and leptodermous idealization, the initiation of sliding and  rolling like motions and the sticking configuration in a dinucleus  were studied in the seminal work by Tsang\cite{tsa74}, where, instead of outgoing fission fragments as in our case, a pair of incoming colliding nuclei of mid-mass  that  experience  frictional forces was considered.  The initial phase of the nuclear collision has weak damping characteristics wherein the sliding and rolling motions arise. In the center of mass frame, once the projectile and target  approach each other sufficiently closely with a finite impact parameter, they initially slide on each other and experience sliding friction. Frictional force will be maximum when the relative velocity at each point in the overlapping region is the highest. As a result of  friction, the relative velocities at these points decrease and when the  relative velocity at the contact point is zero, the sliding motion would stop. Rolling motion will  continue if  the relative angular velocity is non-zero. By the action of rolling friction, which is lower in magnitude, the  relative angular velocity also will decrease to zero. The latter condition causes  the dinuclear system to get stuck completely and rotate as a  rigid rotor. The sliding frictional force transfers the initial orbital angular momentum $L$ into colliding partner individual  angular momenta $I_1$ and $I_2$ which lead the two partners to roll on each other.  By the action of rolling friction, the $L$ transfer gets completed and  the  partners get stuck and  become a rigid rotor.  Equations of motion for the radial direction, individual angular momenta $I_1$ and $I_2$, and the orbital angular momentum $L$ have been obtained with a single frictional coefficient (K) as the only parameter in \cite{tsa74,vol78}. Assuming that the mass asymmetry degree is frozen, and with $\rho_1$ and $\rho_2$ as densities of first and second nuclei in the overlapping region, the equations of motion (except the radial) are given by

\begin{equation}
\dot{I_1} = K {\int} {\rho_{1}} {\rho_{2}} {d^3}{\tau} [(\dot{\theta}R -\dot{\theta_1}R^{'}_{1} -      \dot{\theta_2} R^{'}_{2})R^{'}_{1} - {g^2}( \dot{\theta_1}-\dot{\theta_2})]  ,
\end{equation}

\begin{equation}
\dot{I_2} = K {\int} \rho_{1} \rho_{2} {d^3} \tau [(\dot{\theta}R -\dot{\theta_1}R^{'}_{1} -      \dot{\theta_2} R^{'}_{2})R^{'}_{2} + {g^2}( \dot{\theta_1}-\dot{\theta_2})] ,  
\end{equation}
and
\begin{equation}
\dot{L} = - (\dot{I_1} + \dot{I_2})  .
\end{equation}

As dynamical variables, the polar coordinates of nuclear centers $R$ and $\theta$, and the angles $\theta_{1}$ and $\theta_{2}$ that characterize the space orientation of both nuclei, are used in the above equations. Dots over the symbols denote time derivatives. The arm lengths $R^{'}_{1}$ and $R^{'}_{2}$ are the distances from the nuclear centers to the centroid of the overlap region and they are also measures of the respective nuclear deformations.  Parameter $g$ is the effective radius of gyration of the overlap region around its centroid. 

Equation (3) implies that if $I_1$ and $I_2$ are increased during the collision process due to friction, then $L$ is decreased and vice versa, and it ensures conservation of the total angular momentum. The equations of motion for  $\dot{I_1}$ and $\dot{I_2}$ ( Eq.(1) and Eq.(2)), each has two terms in the square bracket, of which the first term is the major one. This major  term has in the parenthesis the relative tangential velocity of the two colliding nuclei in the over lapping region. The relative tangential velocity gives rise to the sliding frictional  force which when multiplied by the respective arm lengths, $R^{'}_{1}$ and $R^{'}_{2}$,  cause a torque pair. These torques cause non-zero time-derivatives of the same sign for individual angular momenta $I_1$ and $I_2$ but their values  could be  different  due to differences in the arm lengths. Whereas the smaller second term is related to the relative angular velocity of the pair $( \dot{\theta_1}- \dot{\theta_2})$ and this term imparts its influence on   $\dot{I_1}$ and $\dot{I_2}$   of equal magnitude but opposite in sign.

In more sophisticated microscopic models the energy, angular momentum, and nucleon exchange between the nuclei are considered as transport or diffusion phenomenon on the basis of single particle model of the nuclei. However the essential average properties for the interaction of the two nuclear systems are provided by the macroscopic dynamical model given in Ref.\cite{tsa74}. 

\subsection{Similarities of rotational mode environments in fission and heavy ion collisions }
\label{corr}
As mentioned above in Sect.\ref{irmh}, in the entrance phase of  heavy ion collisions, the partners first undergo  sliding motion and then undergo  rolling motion. Whereas the fission reaction, while it proceeds towards the scission stage, has similarities to the entrance phase of heavy ion collisions but in a time reversed manner. The  prefragments, in their journey to scission, first undergo  rolling motion and then reach the stage of nucleon pair transfers from the neck. The 'pair transfers from neck' are  like the sliding motion; they cause 'local' high relative tangential velocity for the participating nucleons. 

The rolling frictional motion causes  $\dot{I_1}$ and $\dot{I_2}$ of same magnitude but opposite in sign. This is a characteristic of  occurrence of the bending mode. The frictional motion generated due to the relative tangential velocity (arising due to pair transfers) will have  $\dot{I_1}$ and $\dot{I_2}$ of  same sign but possibly of different magnitudes in the two prefragments due to their different arm lengths. The pattern of motion generated  due to pair transfers therefore has characteristic of  the wriggling mode. The wriggling mode occurs much more strongly than the bending mode\cite{tsa74,dos85}.  The net result of these actions can lead to probable unequal values for $\dot{I_1}$ and $\dot{I_2}$, at   same time satisfy the condition given by Eq.(3). Recent advanced experimental and theoretical works have renewed interest in this topic and they forwarded competing explanations\cite{wil21,ran21,bul22}, however we maintain that, as also mentioned earlier, relevant average behaviors  are explained by the macroscopic dynamical model given originally in\cite{tsa74}. 

The pair transfers  can also have off-axis displacements due to which the twisting mode is generated. The twisting mode populates angular momentum components along the dinuclear symmetry axis but in opposite directions in the two prefragments so that no net angular momentum is generated in the dinucleus as a whole. The axial rotation (tilting) mode  is not directly excitable by the pair transfers, for which the activation is due to the orbital rigid-rotation present in the dinucleus\cite{ran85}. Note that the  tilting confers angular momentum component along the  symmetry axis, and only if the total angular momentum $I$ of the dinucleus is non-zero. The relaxation time for the tilting mode is relatively long compared to other intrinsic rotational modes owing to its different origin.

\subsection{Nonequilibrium relaxation}
\label{nre}
The mechanism of the intrinsic rotational modes is described above.  How the available free energy fuel these modes,  from among multitude of many other degrees of freedom?  It  depends on how relevant flows and forces  couple among themselves and exchange the energy available. For such processes, the formulation of non-equilibrium sub-system thermodynamics in terms of the free energy dissipation is very useful. This is particularly so  when detailed knowledge of the flows and forces are non-existent. The formulation we used relates firmly the forces and fluxes to a free energy difference in terms of the relative entropy\cite{qi01}.

Relatively slow relaxation of  rotational modes as compared to the nucleonic  degrees and still slower relaxation of elongation(fission) degree allows one to calculate  temperature $T$ of  the heat bath at various instances of elongation. As the nucleon transfers  ceases at scission elongation, $T$ reaches its saturation value equal to $T_s$ thereby  marking the end of the transient phase of thermal generation of intrinsic rotational modes. Till the transient phase ends the dissipative part of the dynamics cause the system to experience a monotonic change of the relative entropy\cite{ao08} as shown in Sect.~\ref{res} below.

\subsection{Statistical model for density of states at a given $T$}
\label{smt}

The wriggling, bending, twisting and tilting mode populations are assumed to  obey statistical distributions with respect to their population probabilities and relaxation times as they are embedded in a heat bath of temperature $T$. The total angular momentum $j$ in a prefragment is the sum of  contributions from individual particles, conferred by the intrinsic rotational modes as well as the rigid rotational or single particle degrees. In thermal occupation condition, the density of states of a given total angular momentum $j$ in a prefragment has been obtained as
\begin{equation}
P(j) = \frac{( 2j+1) }{2\sigma^2}\exp( -\frac{j( j+1) }{2\sigma^2})
\end{equation}
where $\sigma^2$ is the ‘spin-cutoff’ parameter given by  $\sigma ^{2}=\frac{\mathcal{I}T}{\hbar^{2}}$  where $\mathcal{I}$  is the moment of inertia and $T$ is the temperature\cite{eri60}. This form for fragment angular momentum distribution has been experimentally validated in numerous studies, see for instance\cite{wil21}. How the axial alignment of the prefragments modify the density of states? Thermal excitation of the twisting mode results in a certain width for the intrinsic spin states that occur parallel to the dinuclear symmetry axis. Summation over these intrinsic spin states  leads to same above form for $P(j)$ even when the prefragments  are axially aligned\cite{joh98}.

Time dependent values of $P(j,t)$ and steady state saturation value of   $P_{ss} \left( j \right)$  are calculated using relevant values of the temperature and  moment of inertia. The distribution  $P_{ss} \left( j \right)$  has to be the same with and without the knowledge of the damping route the nucleus has taken till scission, whereas $P( j,t)$, on contrary,  depends on  the time scales of excitation energy and matter buildup in the prefragment. (The latter aspect is discussed  in Sect.~ \ref{fra}.)

\subsection{Generalized non-equilibrium free energy and entropy balance equation}
\label{gen}
The transient nonequilibrium deviations of the subsystem  can be described by a generalized free energy difference ($F(t)$) associated with this process\cite{qi01}. In this exploratory work using $F(t)$, it is assumed that the prefragments are of symmetric mass and the shell structure effects are not taken into consideration.  $F(t)$ in one of the forming prefragments is given by
\begin{equation}
F \left( t \right) =~ k_{B}T(t)~ \sum _{j}^{}P \left( j,t \right) \ln \frac{P \left( j,t \right) }{P_{ss} \left( j \right) }   ,
\end{equation}
where  $P(j,t)$ is the probability distribution for total angular momentum $j$ at an instant of time $t$ and $P_{ss}(j)$  is the  equilibrium or steady  probability distribution for $j$ reached at the end of the transient phase. $T(t)$ is the absolute temperature and $k_B$ is the Boltzmann constant. How an arbitrary distribution$P(j,t) \neq P_{ss}(j)$ is changing with time and approaching $P_{ss}(j)$ is determined by  $F(t)$. 

$F(t)$, which is also known as relative entropy, is the free energy difference between a distribution at a point of time $t$ on way to equilibrium, and the distribution at equilibrium \cite{qi01}. It is well known that  fluctuations at the equilibrium returns to its mean value driven by a thermodynamic force. The same thermodynamic force drives the non-equilibrium relaxation towards the equilibrium distribution. The equilibrium fluctuations and non-equilibrium relaxations are thus connected\cite{qi01,maz99}. The connection is given formally by $F(t)$.  It has many important mathematical properties, for example, it is positive and it always decreases until it reaches its minimum at equilibrium. 

The free energy dissipation rate $f_d$ is given by
\begin{equation}
f_{d}=-\frac{dF}{dt}=Te_{p} 
\end{equation}
where $e_p$ is the rate of entropy production and $T$ is the   absolute temperature. In  literature $e_p$ is referred also as non-adiabatic entropy production rate and it is purely determined by the damping mechanism\cite{qi14}. 

The entropy of the subsystem of rotational modes increases due to the entropy generated in spontaneous irreversible processes and decreases when heat is expelled into the surrounding heat bath. It is has been shown\cite{qi14,pri67,ge10} that for such a system, the net rate of change of entropy as a function of time is given by the well known entropy balance equation as

\begin{equation}
\frac{dS \left( t \right) }{dt}= e_{p} \left( t \right) -e_{e} \left( t \right)  
\end{equation}

where $e_p$  is the rate of entropy production in the subsystem and $e_e$ is the rate of entropy expulsion to the surrounding heat bath and  both are functions of time. Using  the methods of non-equilibrium thermodynamics the  rate of entropy production $e_p$ is calculated from Eq.(6). The rate of entropy expulsion $e_e$ is given by $e_e=h_d⁄T$, where $h_d$ is the heat dissipation.

\subsection{Fragment excitation buildup during diabatic phase }
\label{fra}
In non-equilibrium thermodynamics for nuclear fission,  dissipation and fluctuation are the key ingredients. The transient phenomenon of  excitation energy  buildup in the dinucleus ($preTXE$) leads to the heat dissipation in  prefragments given by $h_d$. A simple formalism has been used for calculation of $preTXE$. We  presently adopt simplified approaches such as this with the expectation that  it will be be sufficient for a first orientation towards the main aim of this work, i.e., the first-time application of entropy balance equation  in depiction of the intrinsic rotational modes as dissipative structures. 

It is known that an excited nucleus de-excites itself  and become relatively cold early in the fission process  by emission of particles.  Therefore it is assumed that a net longitudinal motion in the  elongation coordinate $R$ gets initiated from  a relatively cold nucleus.  At an early stage like the saddle point, the deformation forces and  consequential  net  longitudinal motions are practically zero, and the motion  picks up only  during the saddle to scission period. A reduced elongation parameter $\bar{R} = R-R_s$ is defined, where, $R_s$ is the value of elongation parameter at the saddle point. Equation of motion for the reduced elongation parameter $\bar{R}$ is given by
\begin{equation}       
M_{eff}\frac{d^{2}\bar{R}}{dt^{2}}=k\bar{R}  
\end{equation} 
where  $\bar{R}$  vary within the bounds of saddle point and  scission point values and $k$ is the constant of proportionality with positive sign \cite{wal88}.The effective mass   $M_{eff}$  is approximated to be a constant value given by the  reduced mass of the binary nascent  fission fragments. This is a good approximation within the limits of weak to non-damping situations \cite{swi72}. The solution for the equation of motion with initial condition $\bar{R} \textsubscript{(t=0)} = 0$ is given by \cite{wal88}
\begin{equation}
\bar{R}=c \sinh  \lambda t 
\end{equation}
which leads to 
\begin{equation}
\dot{\bar{R}}=c \lambda  \cosh  \lambda t 
\end{equation}
with constants $c$ and  $ \lambda $ determined from empirical considerations and the constant of proportionality is given by
\begin{equation}
k= \lambda ^{2}M_{eff}~.
\end{equation}
The kinetic energy from the above one dimensional model of motion is given by   $E_{K}=\frac{1}{2}M_{eff}~\dot{\bar{R}}^{2}$.  In this  model of saddle to scission motion,  the transverse motions are not taken into account. There are degrees of freedom with transverse components in the dinucleus; for example, the thermally generated rotational degrees and such degrees have to be taken into account.  Once the restriction to just one dimension is lifted, numerous exchanges occur that result in coherent excitations and  non-coherent intrinsic excitations.  The non-coherent intrinsic excitations together constitute a heat bath of energy content $preTXE$.  An approximation is made in this work that the heat bath acquires half of the quantity of the kinetic energy gain  from   the one-dimensional model and the rest half is retained by the collective motions, i.e., $preTXE = \frac{1}{2} E_K$. Sharing of liberated energy between intrinsic and collective motions is one of the  main characteristic of the weak damping phase \cite{swi72}. On reaching scission, the accumulation of $preTXE$ comes to an end. The reduced elongation parameter  $\bar{R}$ and the  $preTXE$,  both as  functions of time, obtained for  fission of \textsuperscript{236}U nucleus using the above model, are given in Fig.1. Values  of the empirical constants used are $c$ = 0.007 fm and $ \lambda $ = 1.0x10\textsuperscript{-21} s\textsuperscript{-1} \cite{wal88}.   The results presented in Fig.1  are consistent  with the microscopic calculations and empirical observations cited in \cite{you11,sim14,qia21}. It is to be noted that the collective kinetic motions, especially when they are near the originating stage and of small amplitudes, are themselves result of a gather of guided intrinsic motions arising from high dissipation. When collective kinetic motion increases   there will be more coherency built and the dinucleus  reach the weak damping phase.  

For the present model for thermal excitation of rotational modes, the above estimation of  $preTXE$ and its time dependence provide useful inputs. For symmetric fission, each prefragment will have $half$ of  the $preTXE$ as its excitation energy at a given instant.  From this excitation energy, the temperature $T(t)$  is calculated using the value of level density parameter $a$ as $a =A_F$/8 where $A_F$ is the mass of the symmetric prefragment. For n\textit{\textsubscript{th}}+\textsuperscript{235}U fission, a value of scission temperature equal to 1 MeV has been found to provide a good description of the mass yield distributions\cite{lem15}, therefore, this exploratory work used the saturation temperature $T_s$=1 MeV. Reaching this temperature has been used as the criterion that the system has reached  the final scission point.The value of $j_{rms}$ calculated using $T$= 1 MeV in Eq.(4) is 8.6$\hbar$ for the symmetric split.   Experimental systematics of $j_{rms}$ in n\textit{\textsubscript{th}}+\textsuperscript{235}U  fission given in \cite{bec13}  compare well with this estimate. For general applications, a better definition of scission point is surely desirable. 

\section{Results}
\label{res}

\begin{figure}
\begin{center}
	\includegraphics[width=80mm,height=60mm]{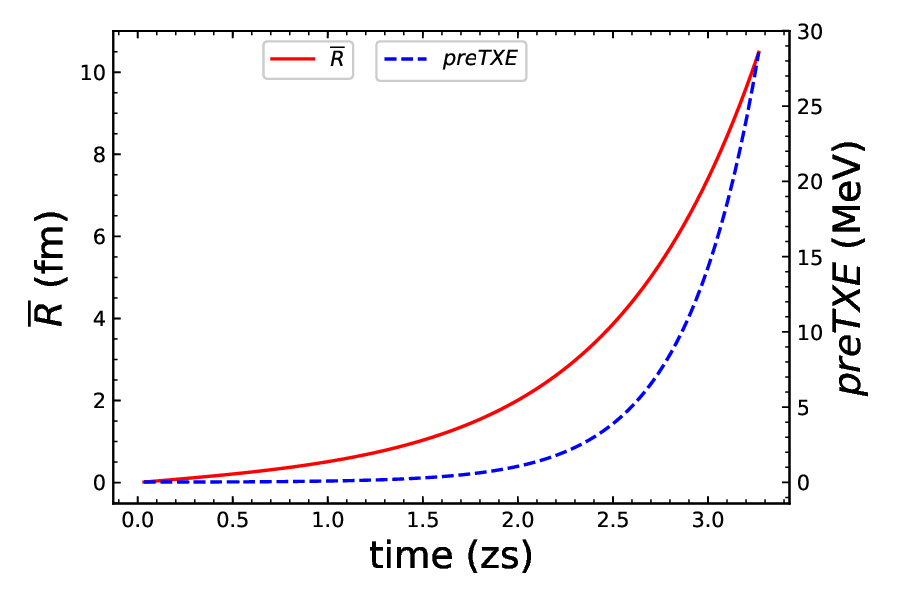}
	\caption{Time evolution of  the reduced elongation parameter $\bar{R}$ (left ordinate) and the excitation energy build-up \textit{preTXE} (right ordinate) during saddle to scission motion in \textsuperscript{236}U fission. }
\end{center}
\end{figure} 
\begin{figure}
\begin{center}
	\includegraphics[width=80mm,height=60mm]{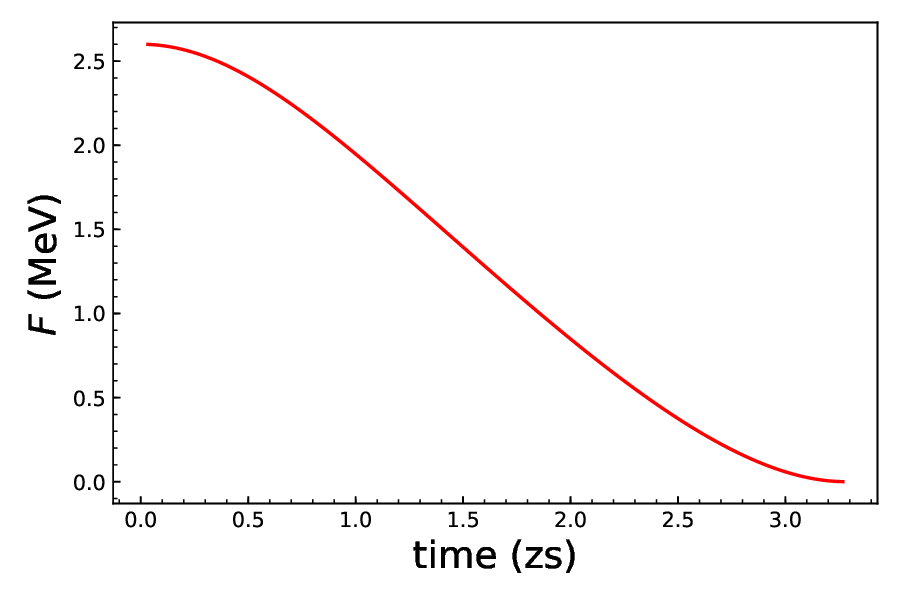}
	\caption{Time evolution of the generalized non-equilibrium free energy difference $F$ relevant to thermal excitations of  rotational modes  in \textsuperscript{236}U fission. }
\end{center}
\end{figure} 
\subsection{Free energy and entropy rates}
\label{fre}
The generalized non-equilibrium free energy difference $F$ and  the entropy production rate $e_p$ are calculated using Eq.(5) and Eq.(6), respectively. The free energy difference $F$ is the thermodynamic potential difference for the process and its decreasing rate is the entropy production rate $e_p$ times the temperature $T$. The free energy difference $F$ as a function of time  is plotted in Fig.2 for low energy fission of  \textsuperscript{236}U.
From the rate of change of prefragment excitation energy, the associated value of heat dissipation $h_d$ is calculated and the corresponding entropy expulsion rate is given by $e_e=h_d/T$.The rates $e_p$ and $e_e$, and the rate of change of entropy  $dS/dt$(Eq.(7)),  are  plotted as a function of time in Fig.3 using dashed, dash-dotted, and continuous lines, respectively. The value of $e_e$ decreases to zero from its value at scission as the prescission heat dissipation ends at scission. Since energy and entropy are extensive variables,  the rates of entropy production and expulsion are additive so that the contributions from both  fragments can be added when necessary. The generalized non-equilibrium free energy $F(t)$ is for a finite volume $V$ where $V$ is also an extensive variable, and  the rates  $e_p$ , $e_e$, and $dS/dt$ are also functions of the volume $V$.
\begin{figure}
\begin{center}
	\includegraphics[width=80mm,height=60mm]{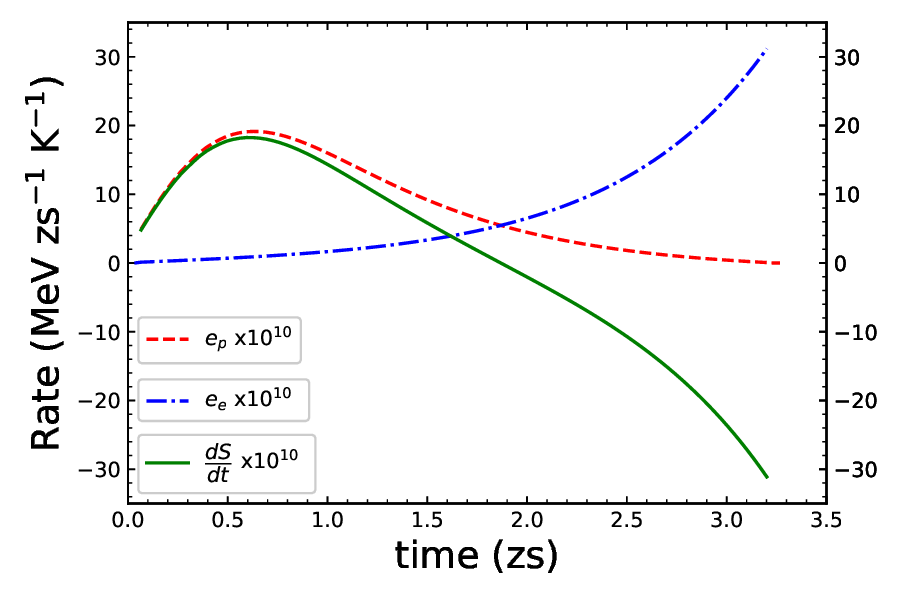}
	\caption{Entropy production rate $e_p$, entropy expulsion rate $e_e$, and  the rate of change of entropy $dS(t))/dt$ as a function of time, relevant  to thermal excitation of  rotational modes in \textsuperscript{236}U fission.The calculated quantities have been multiplied by 10\textsuperscript{10} for display purpose. }
\end{center}
\end{figure} 
\subsection{Emergence of rotational modes}
\label{col}
The  irreversible processes  in  the  dinuclear rotation are described by  Eq.(5) to Eq.(7) in terms of the non-equilibrium thermodynamics. As the nucleon pairs transfer their kinetic energy to the prefragments, a part of the energy, with a spread from fluctuations in its values, is utilized to build up the subsystem of dinuclear rotations. Initially, as many associated  degrees of freedom with the above spread in energy  get opened up, the entropy production rate $e_p$  would increase. From the perspective of  information entropy, the increase in $e_p$ comes about by discarding information due to the spreading out and from information loss associated with the couplings. As the rotations are limited by an energy budget due to the spin-cutoff values, the initial  increase of $e_p$ is followed by a decrease  but it remains a positive quantity ($e_p\geq 0$). The  positivity  of entropy production rate $e_p$ is a general criterion of irreversibility of the process. The part of the nucleon kinetic energy that is unutilized for rotations is assumed to cross the boundary of the ‘open subsystem of rotation’ and it gets dissipated in the heat bath. Therefore the term $e_e$ carry negative sign in the entropy balance equation. The resultant rate of change of entropy $dS/dt$ (Fig.3  continuous line) confirms that  an efficient free energy dissipation  and associated transfer of a higher amount of  lower-grade heat across the boundary can change its sign from positive to negative. This change results in spontaneous high degree of organizations in space, time, and function, and it ultimately  lead to the emergence of intrinsic rotational modes. This  is  a novel description for the intrinsic rotational modes. Under appropriate non-equilibrium conditions, a small fraction of the microscopic motions is organised into a well defined macroscopic motion such as a rotational mode exemplifying  a dissipative structure. 

The entropy rates have the form of power per Kelvin. While the free energy $F$ drops gradually, two regimes are displayed; first an 'incoherent' power transfer followed by a more coherent power transfer. During this process of power transfer, a fraction of the power may get reflected from the dinucleus in the form of radiation. Tracking the radiated power using advanced radiation detector setups may provide  experimental confirmations for these important quantities.

\section{ Discussion}
\label{dis}
The importance of infusing   concepts associated with diabaticity, open subsystems, free energy, and spontaneous emergence of order in dinuclear systems as discussed in this work lies in obtaining a  better understanding of  dissipative systems like the fissioning nucleus. For instance, the free energy dissipation rate $f_d$ (Eq.(6)) signifies an instantaneous current of  the thermodynamic process. The dissipative systems export entropy to its environment to maintain a low internal entropy for itself. As the above mechanism evolves, the system's 'metabolic' paths become more efficient. The mechanism of scission itself, a random rupture or a less or more violent process,  will be guided by the system's  dissipative paths. Decades-old gray area such as the scission process description needs to be looked at from newer perspectives and the present work brings in one. Moreover, the literature on thermally excited intrinsic rotational modes in fission often cites an elusive and abstract nature for them. This is partially due to lack of information on how these modes are initiated by the nucleon motions and their growth in magnitude as  collective modes. Using a simple formalism, the present work brings a clearer physical picture of the initiation and growth of the intrinsic rotational modes, particularly the dominant wriggling mode. The wriggling mode is excited preferentially in a dinucleus and this aspect is already noted in the study of damped nuclear reactions \cite{dos85,ranpl82}.   

Out of two types of non-equilibrium processes widely studied, namely transient and stationary, the emergence of rotational modes is of transient nature. Whereas, the second type, the stationary non-equilibrium process,  is a driven process  and this type requires  sustained energy supply. For all such non-equilibrium systems, the free energy function assumes a leading role, and the entropy production in a non-equilibrium system is regarded as a matter of primary importance. The interest is not only focused on why the entropy change  as the system evolves, but also on how the entropy is produced. The reason being that the entropy production rate is  determined purely by the damping mechanism and hence the irreversibility. The present work probes some of these important aspects in context of nuclear fission, where a dramatic rearrangement of nuclear matter occur with accompanying large energy release. Far from equilibrium nature  is a characteristic  of many other finite many-particle quantum  systems as well, such as atoms and atomic clusters, and similar systems in quantum chemistry and  biology. In comparison to these systems, the transition energies in nuclear fission are far larger and this gives advantages to dedicated experimental investigations where results obtained can be connected to the  free energy functional  and  the entropy production rate. 

\section{ Summary and conclusions}  
\label{sum}
In summary, we considered the diabatic evolution of pre-scission nucleus in a duration  when  the dinuclear intrinsic rotational degrees of freedom emerge in to existence, by using certain thermodynamic concepts that normally characterise macroscopic systems. As is well known, many properties of nuclei are indeed successfully explained using macroscopic formalisms and nuclei are ‘macroscopic objects’ owing to that they contain many particles and a very large number of states. Rotational modes in fission nuclei that cause ample gamma ray emissions  remained barely understood even by the macroscopic formalisms. We analysed the issue of intrinsic rotational modes  from a new perspective in terms of  thermodynamics of irreversible processes. A reasonable explanation has been given by identifying the intrinsic rotational modes as an open subsystem that exchange matter and  energy by means of  coupling to a heat bath that   comprises  of the balance degrees of freedom of the system. Essentially,  the available energy has to be appropriately coupled to the work needed for emergence of self-ordering rotational modes. As the initiator of intrinsic rotational modes,  the pair transfer of nucleons from neck region play the role of such a coupling. Justification for this identification is provided in part by the equations of motion derived under certain macroscopic idealizations \cite{tsa74}. A simple formalism is presented such that the magnitude of relevant energy flows through the subsystem  are  obtained in terms of entropy production rate and entropy expulsion rate in sub-units  of MeV~zs\textsuperscript{-1}~K\textsuperscript{-1}. The rate of change of entropy $dS/dt$ which is the difference of the above rates show a transition from positive to negative values as the dinucleus reached the scission point. This signifies emergence of a new order at a higher level of organisation for the subsystem, i.e., the emergence of intrinsic rotational modes as  dissipative structures. 

\end{document}